# An Electrochemical Technique for Measurements of Electrical Conductivity of Aqueous Electrolytes

**Rajkumar S. Patil, Vinay A. Juvekar* and Umesh Nalage**
Department of Chemical Engineering
Indian Institute of Technology Bombay, Powai, Mumbai-400076, India

**Abstract**

The technique presented here for the measurement of electrical conductivity is based on the principle that the current converges on a small disk electrode. Most of the ohmic resistance therefore lies within a narrow region surrounding the disk. If the reference electrode is kept outside this zone, the potential difference between the working and the reference electrode includes practically all ohmic potential drops occurring in the solution. Moreover, this ohmic drop can be related to the conductivity of the solution by an analytical expression derived by Newman. At sufficiently high overpotentials, the rate of charge transfer is limited by the conduction of current from the bulk solution to the electrode. In this regime, the current varies linearly with the electrode potential and the conductivity of the solution can be estimated from the slope of the voltammogram using Newman's expression. The electrochemical reaction used for measuring conductivity of solutions of salts is the cathodic reduction of water and that used for aqueous acids is the cathodic reduction of hydrogen ions. The technique has been used to measure conductivity of several common aqueous electrolytes. A good agreement is found between the present technique and the conventional technique based on AC impedance analysis.

*To whom correspondence should be addressed. Tel: +91-22-25767236, Fax: +91-22-25766895, E-mail: vaj@iitb.ac.in

# 1 Introduction:

Electrical conductivity is an important property of electrolyte solution. It is a useful tool for monitoring salinity[2], ionic strength[3-5], major solute concentration[6,7], and concentration of the total ionizable solids in water[8-10]. Electrochemists often use conductivity data to gain insight into the structure of electrolyte solutions. Electrical conductivity is also one of the most important parameters in process design and control of electrochemical reactors[11].

In principle, electrical conductivity of a solution can be estimated from the measurement of the potential drop across a DC cell as a function of the current density. However, in this case, the potential drop is not purely ohmic. For the direct current to flow through the cell, electrochemical reactions must occur at the electrodes. The additional potential required to drive these reactions cannot be accurately estimated. In order to eliminate the reaction potential, Gunning et al.[12] have proposed a technique, which employs a cell having a pair of reversible electrodes, both of which are based on the same redox reaction. In this case, the potential drop at the cathode exactly negates that at the anode, so that the measured potential drop is entirely ohmic. However, this technique is of limited applicability owing to the difficulty in preparing reversible electrodes having stable and reproducible electrode potentials.

An alternative and well accepted technique is based on the use of alternating potential, where the measurement of the double layer charging current is used to estimate the conductivity. The advantage of the technique is that no electrochemical reaction needs to occur at the electrodes and hence measurements can be conducted at cell potentials lower than 1V there by completely eliminating the interference from electrochemical reactions. However, the total impedance of the AC circuit is not purely ohmic, but also includes capacitive reactance associated with the double layer. Since the latter decreases with increase in the frequency, conductivity meters use high AC (approximately 2 to 5 kHz) frequencies to minimize the capacitive reactance of the solution. With increasing concentration of the electrolyte, the double layer capacitance increases and hence increasingly higher frequencies are required to suppress the capacitive contribution. At high frequencies, the contribution from stray capacitance becomes significant. More complex circuits are needed for accurate measurement of conductivity in these cases.

In this paper, we describe a novel electrochemical technique for measurement of electrical conductivity. It uses a platinum rotating disk electrode. The principle of the technique is as follows. Current converges on a small disk electrode. This results in a high current density in the close proximity of the electrode. Since the rate of electrochemical reaction increases exponentially with the overpotential, at sufficiently high overpotentials, the rate of charge transfer is limited by the conduction of current from the bulk solution to the electrode. Due to the convergence of the current, most of the ohmic resistance lies within a narrow region surrounding the disk. If the reference electrode is kept outside this zone, the potential difference between the working and the reference electrode includes practically all ohmic potential-drop occurring in the solution. Moreover, this ohmic drop can be related to the conductivity of the solution by an analytical expression derived by Newman[1]. This expression allows us to measure the conductivity of the solution. In the conduction limiting regime, the current varies linearly with the electrode potential. Hence this regime can be detected using linear sweep voltammetry. The conductivity of the solution can be estimated from the slope of this linear portion of the voltammogram. Rotation of the electrode serves two purposes. First, it prevents concentration polarization in the vicinity of the electrode. Second, by sweeping the solution at high rates past the disk surface, it prevents supersaturation of the gas generated by the gas evolution reaction at the electrode, and thereby prevents evolution of the gas bubbles. In the cases when the evolution of the gas bubbles is inevitable, high disk-speed facilitates rapid disengagement of the bubbles (which may otherwise interfere with the electrode reaction) from the electrode surface,

The most common choice of the electrochemical reaction, for the purpose of conductivity measurements, is hydrogen evolution from water. Water is always present in the test solution and hence this reaction has a universal applicability. However, for aqueous solutions of acids, the hydrogen ion reduction is the reaction of choice.

The present technique has the following advantages over the AC measurement technique. No compensation for the capacitive reactance is needed since the contribution from the double layer charging current is negligible. This simplifies the measurement methodology considerably. Also the technique is found to be applicable over a fairly wide range of conductivities, although we have not explored its full range of application in this work.

The aim of the paper is to describe this technique in some details. Work presented here is explorative in nature and considerable improvements are needed before this technique can be brought to a high level of precision.

The layout of the paper is as follows. We first describe the theory behind this technique. We then describe the experimental setup. In the results and discussion section, we first illustrate the technique with an example. We then describe the effect of various parameters on the accuracy of the measurement. We also compare the values of the conductivity estimated by the present technique with those reported in the literature, for a variety of electrolytes. We also discuss the sources of errors and possible ways to improve the measurement accuracy.

**2 Theory:**

In general, a one-step electrode reaction can be described by the Butler-Volmer equation:

$$i = i_0 [exp(\alpha f \eta) - exp\{-(1-\alpha)f\eta\}] \qquad (1)$$

Where $i$ is the current density at the electrode, $i_0$ is the exchange current density, $\alpha$ is the transfer coefficient, $\eta$ is the overpotential and $f = nF/RT$ (where n, $F$, $R$ and $T$ correspond, respectively, to number of electrons transferred, Faraday constant, universal gas constant and absolute temperature of electrolyte solution).

In practice, the potential of the working electrode is measured with respect to a reference electrode. For the correct measurement of the overpotential, the reference electrode should be placed very close to the working electrode so that the potential difference between the two electrodes does not include the ohmic overpotential. Now consider a situation in which the geometry of the working electrode is such that the current converges on the electrode so that practically all ohmic resistance resides in the close proximity of the electrode. If the potential of the working electrode is measured using a reference electrode which is kept outside the region of current convergence, then the potential difference between the working electrode and the reference electrode will also include the ohmic potential drop and we can write

$$V_e = V_{e0} + iA_e R_s \qquad (2)$$

Here, $V_e$ is potential of the working electrode with respect to the actual reference electrode (which is outside the current-convergence region) and $V_{e0}$ is its potential with respect to a fictitious reference electrode kept so close to the working electrode that it does not sense the ohmic drop. The term $A_e$ is the electrode area and $R_s$ represents the Ohmic resistance of the solution.

At the electrochemical equilibrium, the current density is zero. Hence both the actual and fictitious electrodes will sense the same potential. We denote this potential by $V_{eq}$. Noting that the correct value of the overpotential to be used in the Butler-Volmer equation is $\eta = V_{e0} - V_{eq}$, we conclude from Eq 2 that

$$\eta = V_e - V_{eq} - iA_e R_s \tag{3}$$

Substituting this expression for $\eta$ into Eq 1, we get

$$i = i_0 \left[ \exp\{\alpha f(V_e - V_{eq} - iA_e R_s)\} - \exp\{-(1-\alpha)f(V_e - V_{eq} - iA_e R_s)\} \right] \tag{4}$$

Suppose a linear sweep voltammetry is conducted on this electrode in the anodic direction. At sufficiently high overpotentials, we can reduce Eq 4 to the Tafel form

$$i = i_0 \exp\{\alpha f(V_e - V_{eq} - iA_e R_s)\} \tag{5}$$

Differentiation of Eq 5 with respect to the electrode potential yields the following equation

$$\frac{di}{dV_e} = \alpha f \left( 1 - \frac{di}{dV_e} A_e R_s \right) i_0 \exp\{\alpha f(V_e - V_{eq} - iA_e R_s)\} \tag{6}$$

We first replace the last part of the Eq 6 by $i$ using Eq 5 and then rearrange the equation to yield

$$\frac{di}{dV_e} = \frac{i\alpha f}{1 + i\alpha f A_e R_s} \tag{7}$$

At sufficiently high current density, the term $i\alpha f A_e R_s$ is much greater than unity $(i\alpha f A_e R_s \gg 1)$ and Eq 7 reduces to

$$\frac{di}{dV_e} = \frac{1}{A_e R_s} \tag{8}$$

This equation indicates that, the plot of current density versus potential is a straight line with the slope equal to $1/(A_e R_s)$, from which the solution resistance $R_s$ can be computed.

During the cathodic sweep, Eq 5 should be replaced by

$$i = i_0 \exp\{-(1-\alpha)f(V_e - V_{eq} - iA_e R_s)\} \quad (9)$$

Eq 7 then reduces to

$$\frac{di}{dV_e} = -\frac{i(1-\alpha)f}{1 - i(1-\alpha)fA_e R_s} \quad (10)$$

For sufficiently high cathodic potentials, we have $-i(1-\alpha)fA_e R_s \gg 1$ and Eq 10 reduces to Eq 8. This indicates that Eq 8 is valid for both anodic and cathodic sweeps.

Newman[12] has derived the expression for the potential distribution around a disk electrode, placed in an infinite solution, under conduction limiting regime. The Laplace equation is solved using the rotational elliptic coordinates $(\xi, \eta;\ 0 \leq \xi < \infty, 0 \leq \eta \leq 1)$ to yield the following expression for the potential distribution

$$\frac{V_{e\infty} - V_{e\xi}}{V_{e\infty} - V_{e0}} = 1 - \left(\frac{2}{\pi}\right)\tan^{-1}\xi \quad (11)$$

where $V_{e\xi}$ is the potential of the solution at location $\xi$; The locations $\xi = 0$ and $\xi = \infty$, respectively correspond to the centre of the disk and that which is far away from the disk.

Newman has also related the ohmic resistance of the solution to its conductivity, $\kappa$, and the is the disk-radius $r_d$ by the following equation

$$R_s = \frac{1}{4 r_d \kappa} \quad (12)$$

Combining Eqn 8 and 12, and writing $A_e = \pi r_d^2$ we obtain

$$s = \frac{di}{dV_e} = \frac{4\kappa}{\pi r_d} \qquad (13)$$

Hence the slope of the straight line portion of the linear sweep voltammogram can be directly used to estimate the conductivity of the solution (i.e. $\kappa = s(\pi r_d/4)$).

Eq 13 holds exactly only if the entire resistance of the solution lies between the electrode and the reference probe, i.e. when the reference probe is located at $\xi = \infty$ or where the ratio: $\phi = (V_{e\infty} - V_{e\xi})/(V_{e\infty} - V_{e0})$ equals zero. We call $\phi$ as the fraction of the ohmic resistance of the solution which is not sampled by the reference probe; $\phi = 1$ corresponds to the case where no resistance is sampled (i.e. where the reference probe is kept very close to the working electrode) and $\phi = 0$ corresponds to the case where all resistances is sampled. In Figure 1, we have plotted $\phi$ as function of $\xi$ using Eq 11.

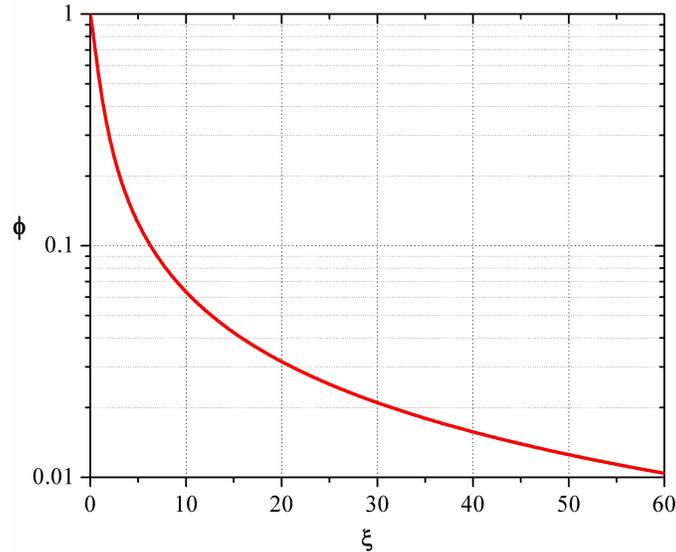

**Figure 1**: Fraction of the total ohmic resistance not sampled by the reference probe as a function of the distance of the reference electrode from the working electrode. The plot is derived using Eq 9.

It is seen that the value of $\xi$ should be about 60 in order that $\phi = 0.01$, i.e. only 0.01 fraction of the Ohmic resistance is not sampled by the probe.

It is important to relate $\xi$ with the distance of the reference probe from the centre of the disk. The rotational elliptic coordinates $(\xi, \eta)$ are related to the cylindrical coordinates $(r, z)$ by the following equations.

$$\frac{r}{r_d} = \sqrt{(1+\xi^2)(1-\eta^2)} \; ; \quad \frac{z}{r_d} = \xi\eta \tag{14}$$

where $r_d$ is the radius of the disk. If $d$ is the distance of the reference electrode from the centre of the disk. Then $d^2 = r^2 + z^2$, and from Eq 14 we obtain

$$\left(\frac{d}{r_d}\right)^2 = \xi^2 + 1 - \eta^2 \tag{15}$$

Since, $0 \leq \eta \leq 1$, $1-\eta^2$ lies between 0 and 1. Thus when, $\xi \gg 1$, we have $d/r_d \approx \xi$. This means that the reference probe should be located at a distance greater than $60 r_d$ from the centre of the disk in order to sample practically all resistance of the solution.

This analysis is based on the premise that the electric field is radially symmetric around the disk. This requires that the counterelectrode is a cylindrical surface of infinite radius and is coaxial with the disk. Such a design is impractical. In the present work, we have arrived at a simpler configuration after several trials. We describe it in the next section. Although, this configuration violates some of the constraints imposed by the theory, we have been able to measure the solution conductivity with acceptable accuracy.

Another important aspect of this technique is the choice of the reaction. Since we have limited the exploration of this technique only to aqueous electrolytes, water is always present in the test solution and hence the reactions which have universal applicability are the cathodic reduction of water

$$H_2O + e^- \rightarrow \frac{1}{2}H_2 + OH^- \quad (E^0 = -0.828V) \qquad (16)$$

and anodic oxidation of water

$$H_2O \rightarrow 2H^+ + \frac{1}{2}O_2 + 2e^- \quad (E^0 = 1.229V) \qquad (17)$$

The advantage of these reactions is that water, which is the reactant, is available in abundance around the electrode and hence its transport to the electrode surface is never limited by diffusion. Between the two reactions, the cathodic reaction is more amenable to measurement since the electrode remains in active state at cathodic potentials whereas it tends to passivate in the anodic region. For this reason, we have utilized reaction (16) for most of the measurements.

For measurement of conductivity of aqueous solutions of acids, it is more convenient to use the hydrogen ion reduction reaction.

$$H^+ + e^- \rightarrow \frac{1}{2}H_2 \quad (\Delta E^0 = 0V) \qquad (18)$$

We have also illustrated its use in this work.

### 3 Materials and Methods:

All experiments were conducted using rotating disk electrode system of Pine Instrument Company, USA. The electrode was a 2.5 mm radius, bright platinum disk embedded in a Teflon sleeve. The rotation speed of the electrode was controlled by a speed controller, and was kept constant at 4000rpm, except in those experiments where the effect of the speed of rotation was studied. The potentiostat used for the measurements had the following specifications: potential range $\pm 10V$, current $\pm 2A$, and the input impedance of the reference electrode of $10^{12}\Omega$.

The cell used for the measurements was 150mm diameter and 50 mm height cylindrical jar as shown in Figure 2. It was filled with the test solution up to a height of 25 mm. The rotating disk working electrode, W, was located at the centre of the cell at a distance of 15 mm from the bottom .The two counterelectrodes (10mm x 5mm rectangular platinum plates), C, were placed equidistant and at diametrically opposite locations with respect to the working electrode. Significant improvement in the accuracy was observed with two symmetric counterelectrodes than with a single (asymmetrically placed) counterelectrode. Saturated calomel electrode (SCE) was used as the reference electrode, R. The best position of the reference electrode was found to be midway between the working and the counterelectrode.

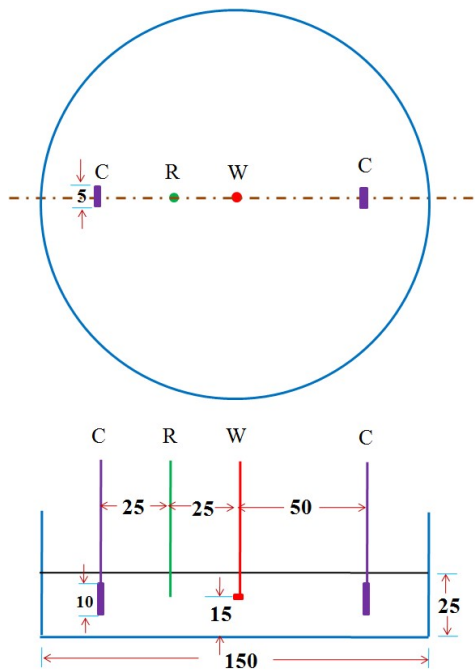

**Figure 2:** Schematic diagram of the cell

All dimensions are in mm. C-Counter electrode, R-reference electrode, and W-working (disc) electrode

All chemicals used in these experiments were analytical grade (Merck, India). The solutions were prepared using Milli-Q water (conductivity, $1.5\,\mu Scm^{-1}$). To maintain inert atmosphere, nitrogen gas was purged in the electrolyte solution for about 20 min before performing the experiment. The platinum disk electrode was polished using polishing kit (ALPHA Micropolish Alumina, Buehler, USA). Conductivity of the test solutions was also measured using Thermo Orion conductivity meter (Model: Orion 145A+, Thermo Electron Corporation, USA). The conductivity meter had the range of conductivity from $0.1\,\mu Scm^{-1}$ to $200\,mScm^{-1}$. The appropriate cells were used in different ranges of concentrations of electrolyte solutions. The cells were precalibrated using the standard KCl solutions supplied by the company.

**4 Results and Discussion:**

*4.1 Illustration of the Technique:*

We illustrate the technique using water reduction reaction in 0.01 *M* aqueous solution of potassium chloride. The solution was subjected to linear sweep voltammetry in the potential range of 0 to -3.5 V and disk speed of 4000 RPM. The sweep rate was kept deliberately high (2 *V.s$^{-1}$*) in order to limit the extent of gas evolution. The resulting voltammogram is as shown in Figure 3.

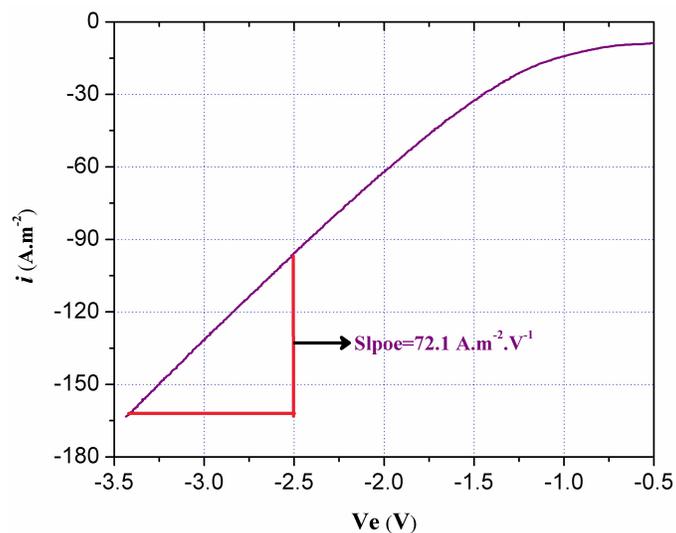

**Figure 3:** Linear sweep voltammogram indicating the conduction limiting regime.

**Test Solution:** 0.01M KCl, Scan rate 2 V.s$^{-1}$, Disk speed 4000 rpm, Temperature 25°C.

It is seen from the figure that beyond the electrode potential of about -2 V, the voltammogram is a straight line. The slope of the line is 72.1 *A.m$^{-2}$.V$^{-1}$*. Using this slope in Eq 13, and the value of $r_d = 2.5 mm$, we estimate the electrical conductivity of 0.142 S.m$^{-1}$. The corresponding value of the conductivity, reported in the literature, is 0.141 S.m$^{-1}$ [13,14].

Next we check whether the potential range from -2.5 to -3.5V is the conduction limiting regime. Based on Eq 7, the condition for the validity of this regime can be written as

$$\left(\frac{\pi \alpha f r_d}{4\kappa}\right) i \gg 1 \tag{19}$$

As seen from Figure 3, the value of the current density at the onset of the conduction limiting regime is about $-100 A.m^{-2}$. Using $\alpha = 0.5$, we compute the value of the left hand side term of the inequality as 26.9, which is indeed much greater than unity.

## 4.2 Effect of the Location of the Reference Electrode:

Location of the reference electrode is of crucial importance for the accuracy of the technique. Based on preliminary experiments, we found that the highest accuracy is achieved when the reference electrode is placed along the line connecting the working electrode and the counterelectrode. Keeping it along this line, we varied the distance of the reference electrode from the working electrode and estimated the conductivity of 0.01 M. KNO$_3$ solution. These estimates are listed in Table 1

**Table 1:** Effect of the distance d of the reference electrode from the working electrode on the measurement of conductivity

**Test solution:** 0.01M KNO$_3$, Temperature: 25$^0$C.

| d (mm) | slope (A.V$^{-1}$.m$^{-2}$) | $\kappa$ (S.m$^{-1}$) |
|---|---|---|
| 10 | 73.82 | 0.145 |
| 20 | 69.75 | 0.137 |
| 25 | 69.75 | 0.137 |
| 30 | 66.70 | 0.131 |
| 40 | 64.15 | 0.126 |

Comparison of the values from Table-1 with that reported in literature (0.136 S.m$^{-1}$ at 25$^0$C[13]) indicates that most accurate estimate of the conductivity is obtained when reference electrode is placed midway between the working electrode and one of the counterelectrodes( note that the distance between the reference and the counterelectrode is 50 mm) . When the reference electrode is brought nearer, the conductivity is overestimated because the reference electrode cannot fully sample the ohmic resistance around the working electrode. On the other hand, if the reference electrode is placed away from the working electrode, that is, nearer he counterelectrode, its potential begins to be affected by the current distribution around the counterelectrode. The midway position is therefore most suitable.

### *4.3 Effect of the Speed of Rotation of the Disc:*

Water reduction reaction is accompanied by the evolution of hydrogen gas. At high cathodic potentials, hydrogen may nucleate at the electrode surface. If the gas bubbles cover the electrode surface, reduction in the effective area of the electrode would result, thereby causing inaccuracy in the measurement. A high speed of rotation of the disk produces high rates of flow of fluid along the disk surface. This results in a rapid transport of hydrogen gas from the electrode surface to the bulk of the fluid, thereby reducing the extent of supersaturation near the electrode. The shear produced by the fluid at the surface of the electrode also helps in disengaging the bubbles that may have nucleated at the surface. Efficiency of the disc is expected to improve with increase in the speed of rotation. To assess this effect, experiments were conducted at different speeds of rotation in the range from 1000 to 5000 RPM. Two test KCl solutions having concentrations of 10mM and 100mM were used. The estimated conductivity value is plotted against the disk speed. The values are compared with those reported in the literature[13].

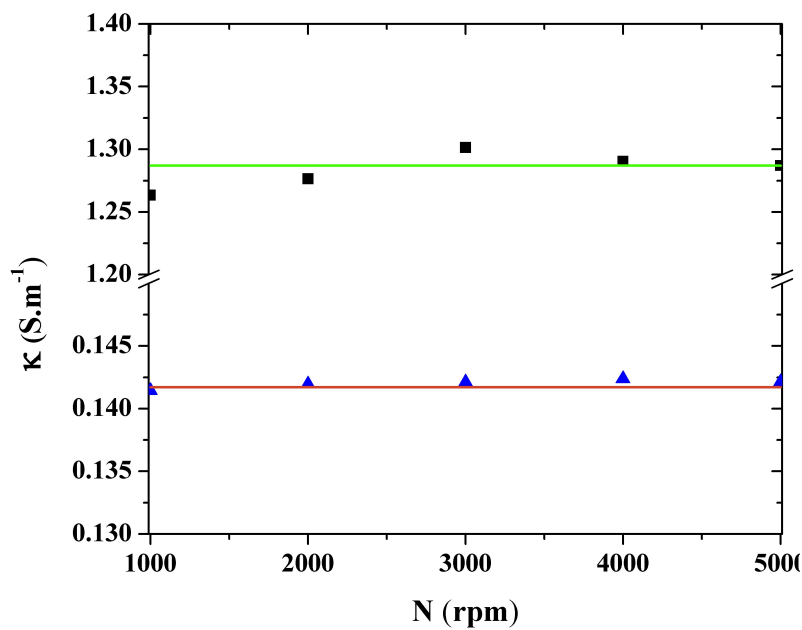

**Figure 4:** Effect of the speed of rotation of disk electrode on measurement of conductivity.

**Parameters:** Scan rate; 2 V.s$^{-1}$, Temperature; 25°C, **Solid triangle**; 10mM KCl, **Solid Square**; 100mM KCl, **Lines** (Values reported in literature[13]); **Red**; 10mM, **Green**; 100mM KCl.

It is seen from the figure that above the disc speed of 2000 RPM, the conductivity measurement is not significantly affected by the disk speed. A slight decrease is observed at or below 2000 RPM. In the present work, all conductivity measurements were carried out at disc speed of 4000 RPM, which is well above 2000 RPM.

*4.4 Effect of the Scan Rate:* High potential scan rates are desirable in order to avoid excessive gas evolution. However, the double layer charging current increases with increase in the scan rate. In order to determine whether the charging current has any significant contribution to make to the total current, the experiments were conducted on the test solution of 10mM KCl at 25$^0$C, by varying the scan rates in the range from 0.3 V.s$^{-1}$ to 4V.s$^{-1}$. The estimated values of the conductivity are listed in Table-2.

**Table 2:** Effect of potential scan rate on measurement of conductivity

**Test solution**: 10mM KCl at 25°C. The value of the conductivity reported in the literature[13] is 0.141 S.m$^{-1}$

| Scan Rate (V.s$^{-1}$) | s (A.V$^{-1}$.m$^{-2}$) | $\kappa$ (S.m$^{-1}$) |
|---|---|---|
| 0.3 | 72.76 | 0.143 |
| 0.5 | 72.07 | 0.142 |
| 1.0 | 72.90 | 0.143 |
| 2.0 | 72.07 | 0.142 |
| 3.0 | 72.02 | 0.142 |
| 4.0 | 72.70 | 0.143 |

It is seen from Table-2 that the scan rate has practically no effect on the measured value of the conductivity. This indicates that the double layer charging current not significant. We verify this observation with approximate calculations. If we assume the double layer capacitance of 0.2 F.m$^{-2}$, the charging current for the scan rate of 2 V.s$^{-1}$ would be 0.2 x 2 = 0.4 A.m$^{-2}$. The current densities used in these experiments are about 100 A.m$^{-2}$. The charging current contribution to the total current is therefore negligible.

*4.5 Effect of Electrolyte Concentration*

As the electrolyte concentration increases, the conductivity increases and hence greater overpotential is needed to achieve the conduction limiting regime. This point is clearly illustrated in Figure 5, which displays the linear sweep voltammograms for three different concentrations of potassium chloride solutions. The location of the point beyond which voltammogram attain linearity is shown by vertical lines on the plot.

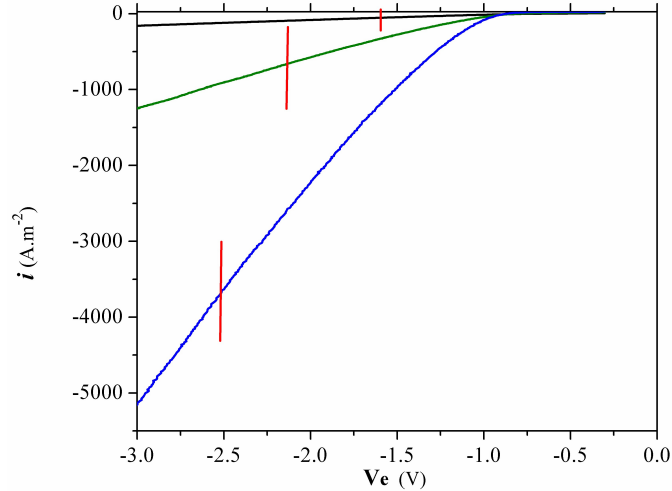

**Figure 5:** Linear Sweep Voltammograms for different concentrations of electrolyte solutions
**Parameters:** Scan rate; 2V.s$^{-1}$, Speed; 4000rpm, Temperature 25$^0$C,

**Lines**: **Black** 10mM KCl (conductivity value; estimated: 0.1415, reported: 0.1413 S.m$^{-1}$), **Green** 100mM KCl (conductivity values; estimated: 1.2803 S.m$^{-1}$, reported: 1.2890 S.m$^{-1}$), **Blue** 500mM KCl (conductivity values; estimated: 5.911S.m$^{-1}$, reported: 5.893S.m$^{-1}$)

We see from the figure that with increasing concentration of the electrolyte, conduction threshold is reached at increasingly higher values of potentials. Coupled with high electrolyte concentration, we expect a high rate of gas evolution at the electrode. This may limit the concentration up to which we can accurately measure the conductivity by this technique.

The results discussed so far are limited to low concentrations of electrolytes. We have also tested this technique at high concentration of electrolytes. For example, we have compared the conductivity of 3M NaCl solution at 25°C by our method (our estimate = $18.4\,S.m^{-1}$) with that reported in the literature ($19.3\,S.m^{-1}$) [13]. The two values are in fairly good agreement.

*4.6 Hydrogen Ion Reduction Reaction*:

Hydrogen ion (proton) reduction reaction dominates over the water reduction reaction when the electrolyte is an aqueous solution of acid. Since the former reaction occurs at a lower (absolute) cathodic potential than water reduction, it has a lower conduction limiting potential. It is therefore more convenient to employ this reaction for the measurement conductivity of both strong and weak acids. This technique is illustrated in Figure 6 which presents a linear sweep voltammogram of 0.01M HCl. The voltammogram is linear beyond -1.0V. The slope of the linear portion is 210 A.m$^{-2}$.V$^{-1}$. The conductivity of the solution is estimated to be 0.412 S.m$^{-1}$ which is in good agreement with the value of 0.4118 S.m$^{-1}$ reported in the literature[15].

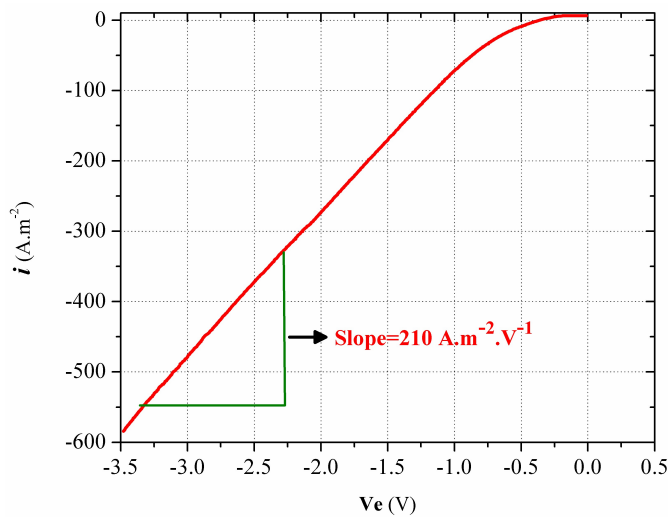

**Figure 6:** Demonstration of hydrogen-ion reduction reaction for conductivity measurement
**Test solution**: 0.01M HCl, disk speed; 4000rpm, scan rate; 2V.s$^{-1}$, Temperature; 25$^0$C

## *4.7 Some Case Studies*:

Using the present technique, we have measured the conductivities of different electrolyte solutions over a range of concentrations. Tables 3 to 5 below list the results. These values are compared with those obtained from the literature. All the reported values are based on AC impedance technique

**Table 3:** Comparison of present technique with AC impedance technique
**Test solution**: Aqueous potassium chloride at 25$^0$C

| Concentration (mM) | $\kappa$ (S.m$^{-1}$) Present Method | $\kappa$ (S.m$^{-1}$) AC impedance[13,14] | (%)Error |
|---|---|---|---|
| 0.5 | 7.64x10$^{-3}$ | 7.39x10$^{-3}$ | 3.27 |
| 1 | 1.47 x$10^{-2}$ | 1.44 x$10^{-2}$ | 2.29 |
| 10 | 0.142 | 0.141 | 0.21 |
| 100 | 1.29 | 1.29 | 0 |
| 500 | 5.91 | 5.89 | 0.30 |
| 1000 | 10.9 | 11.2 | -2.01 |

**Table 4:** Comparison of present technique with AC impedance technique

**Test solution**: Aqueous potassium nitrate at 25$^0$C.

| Concentration (mM) | $\kappa$ (S.m$^{-1}$) Present Method | $\kappa$ (S.m$^{-1}$) AC impedance [13,14] | (%)Error |
|---|---|---|---|
| 0.1 | 1.51x$10^{-3}$ | 1.44x$10^{-3}$ | 4.63 |
| 1 | 1.43 x$10^{-2}$ | 1.42 x$10^{-2}$ | 0.69 |
| 10 | 0.139 | 0.136 | 2.44 |
| 100 | 1.16 | 1.20 | -3.58 |

**Table 5:** Comparison of present technique with AC impedance technique

**Test Solution**: Aqueous hydrochloric acid at 25$^0$C.

| Concentration (mM) | $\kappa$ (S.m$^{-1}$) Present Method | $\kappa$ (S.m$^{-1}$) AC impedance[13,15] | Error (%) |
|---|---|---|---|
| 0.1 | 4.39x$10^{-3}$ | 4.25 x$10^{-3}$ | 3.18 |
| 1 | 4.23 x$10^{-2}$ | 4.21 x$10^{-2}$ | 0.35 |
| 10 | 0.412 | 0.412 | 0 |
| 100 | 3.84 | 3.91 | -1.84 |

It is seen from these tables that our measurements are in good agreement with the reported values except at very low concentrations, where the present method appears to overestimate the conductivity. A possible reason is that there is an additional contribution to conductivity from hydrogen ions (and to a lesser extent from hydroxide ions), which are generated by dissociation of water. This contribution becomes significant only at very low concentrations of electrolyte. This contribution is a strong function of the pH of the solution. We presume that the reported values have been corrected for this extra contribution. Instead of estimating this correction, we have compared, in Table-7, values measured by present method, for these lowest concentration solutions, with those measured by the Thermo-Orion conductivity meter.

**Table-7:** Comparison of the present measurement with those based on AC conductivity meter for very dilute electrolyte solutions at 25⁰C.

| Solution | Concentration (mM) | $\kappa$ Present method (S.m$^{-1}$) | $\kappa$ Thermo-Orion Meter (S.m$^{-1}$) | $\kappa$ Literature[13,14] (S.m$^{-1}$) |
|---|---|---|---|---|
| KCl | 0.5 | 7.64x10$^{-3}$ | 7.57x10$^{-3}$ | 7.39x10$^{-3}$ |
| KNO$_3$ | 0.1 | 1.51x10$^{-3}$ | 1.49x10$^{-3}$ | 1.44x10$^{-3}$ |
| HCl | 0.1 | 4.39x10$^{-3}$ | 4.35x10$^{-3}$ | 4.25 x10$^{-3}$ |

The table shows that the conductivity measured by the Thermo-Orion meter are much closer (within 2%) to those measured by the present technique.

*4.8 Sources of Errors and Possible Improvements*:

The main sources of error associated with this technique are those caused by inadequate sampling of the ohmic resistance of the solution and also the departure from Eq 10 due to lack of radial symmetry in the present cell. Both of these are caused by the improper geometry of the cell, cell-size and the location of the reference electrode and the counterelectrodes. A simple way to improve the accuracy is to calibrate the cell using a standard solution of known conductivity. An appropriate factor can then be introduced in Eq 10 to correct for the errors associated with the cell geometry. Alternatively, one can design a cell which fulfills the necessary conditions for the validity of the Eq 10. For example, one can use a disk microelectrode of say $10\,\mu m$ radius as the working electrode. All the ohmic resistance would then be located within a region of 1 mm radius around the working electrode. It is then possible to use a cylindrical counterelectrode of 10 mm radius, which surrounds the working electrode. The reference microelectrode can then be placed at 5 mm distance from the working electrode. This will allow all the necessary requirements for the accurate measurements to be satisfied. Instead of rotating the electrode, external circulation of the solution around the electrode can be provided in order to prevent accumulation of gas bubbles on the disk surface. It is then possible to use this cell as a primary cell for the measurement of conductivity.

## 5 Conclusion:

In this work, we have demonstrated an electrochemical technique for measurement of electrical conductivity of aqueous solutions. Although, the technique requires further improvements in terms of the system design, we have been able to show that it is simple and yet a versatile. Hence, it has a potential to become an alternative to the presently practiced AC impedance technique.

Authors wish to acknowledge **Unilever India Ltd.** for funding this work.